\documentclass[11pt,a4paper,headinclude,footinclude,chapterprefix=on]{scrreprt}

\usepackage[utf8]{inputenc}
\usepackage[T1]{fontenc}
\usepackage{tabularx}
\usepackage{multicol}
\usepackage{graphicx}

\usepackage{tikz}
\usepackage{lipsum}

\usepackage[headsepline,plainheadsepline,footsepline,plainfootsepline]{scrpage2}

\usepackage{mathptmx}
\usepackage[scaled=.92]{helvet}
\usepackage{courier}

\usepackage{graphicx}
\usepackage[utf8]{inputenc}

\usepackage{amssymb}
\graphicspath{{figures/}}
\usepackage{url}
\usepackage{wrapfig}
\usepackage{units}

\usepackage[printonlyused]{acronym}

\usepackage{calc}
\usepackage{amsmath}
\usepackage{amssymb}
\usepackage{amsfonts}
\usepackage{algorithm}
\usepackage{algpseudocode}
\usepackage{color}
\usepackage[caption=false]{subfig}
\usepackage{stfloats}
\usepackage{nicefrac}

\usepackage{multirow}

\usepackage[multiuser,nomargin,marginclue,footnote]{fixme}
\fxusetheme{color}
\fxsetup{targetlayout=colorcb}
\FXRegisterAuthor{all}{anall}{ALL}
\FXRegisterAuthor{md}{anmd}{MD}
\FXRegisterAuthor{tolja}{antolja}{TOLJA}
\FXRegisterAuthor{pg}{anpg}{PG}
\FXRegisterAuthor{mch}{anmc}{MC}
\FXRegisterAuthor{cp}{ancp}{CP}
\FXRegisterAuthor{awo}{anawo}{AWO}
\fxsetup{status=draft}

\usepackage{geometry}
\clearscrheadfoot

\ihead[\Large {\scshape TU Berlin }]{\Large {\scshape TU Berlin }}

\ifoot[{\tiny \begin{minipage}{4.0cm}Copyright at Technische Universität Berlin.\newline All Rights reserved.\end{minipage}}]{{\tiny \begin{minipage}{4.0cm}Copyright at Technische Universität Berlin.\newline All Rights reserved.\end{minipage}}}
\ofoot[Page \pagemark]{Page \pagemark}

\addtokomafont{pagefoot}{\normalfont}

\newcommand{\trnumber}{TKN-16-002} % please use 3 digits for the report number
\newcommand{\trdate}{July 2016}
\newcommand{\trauthor}{Piotr Gawłowicz, Sven Zehl, Anatolij Zubow and Adam Wolisz}
\newcommand{\tremail}{\{gawlowicz, zehl, zubow, wolisz\}@tkn.tu-berlin.de}
\newcommand{\trtitle}{NxWLAN: \textbf{N}eighborhood e\textbf{X}tensible \textbf{WLAN}}

\begin{document}

%!TEX root = ../techreport.tex

% ================================================================
% Cover Sheet
% ================================================================
{
\sffamily

\thispagestyle{empty}

\setlength{\tabcolsep}{0pt} % remove indent within cell
\noindent % remove indent before cell
\begin{tabularx}{\columnwidth}{cXc}
  \includegraphics[height=1cm]{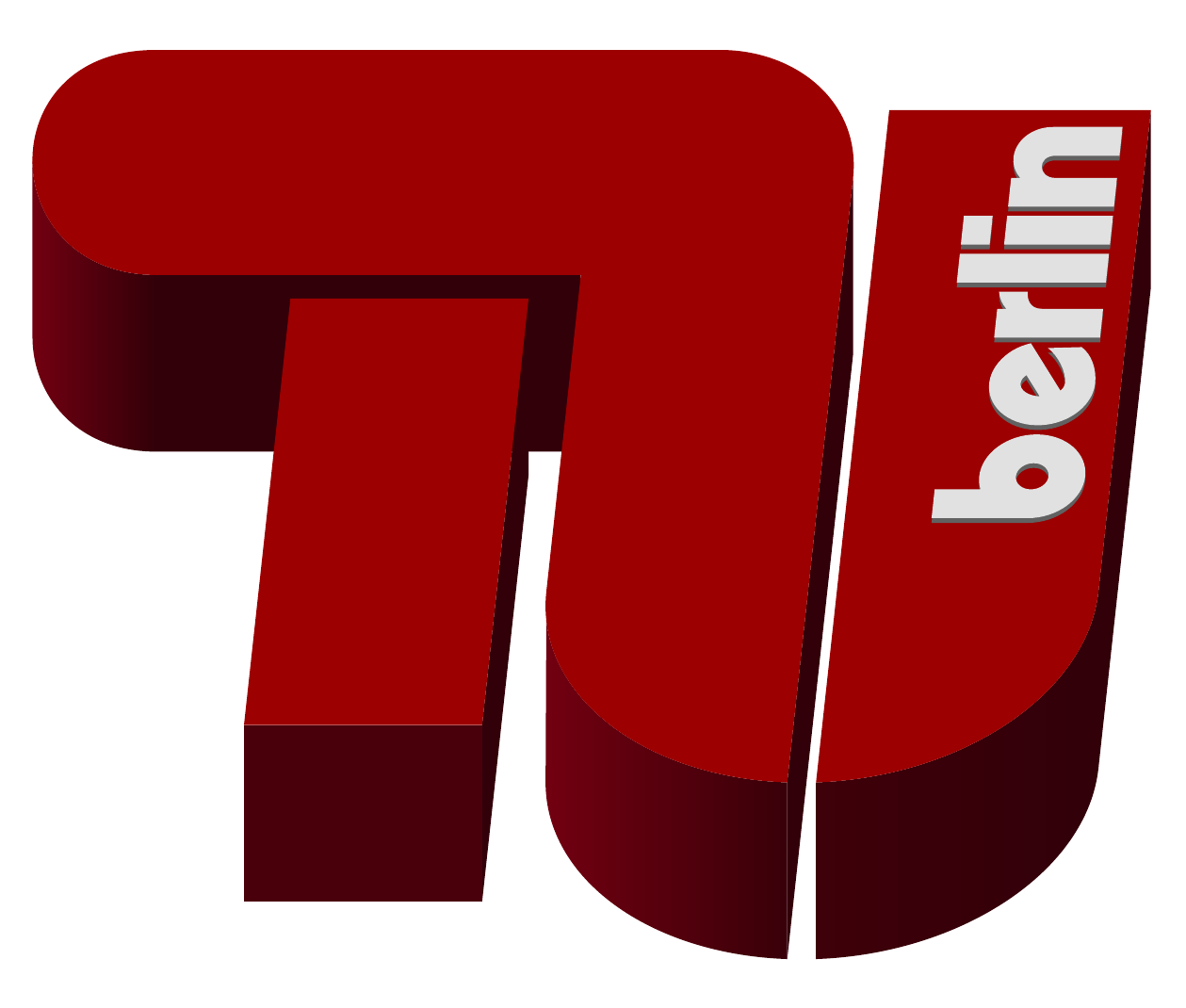}
  & &
  \includegraphics[height=1cm]{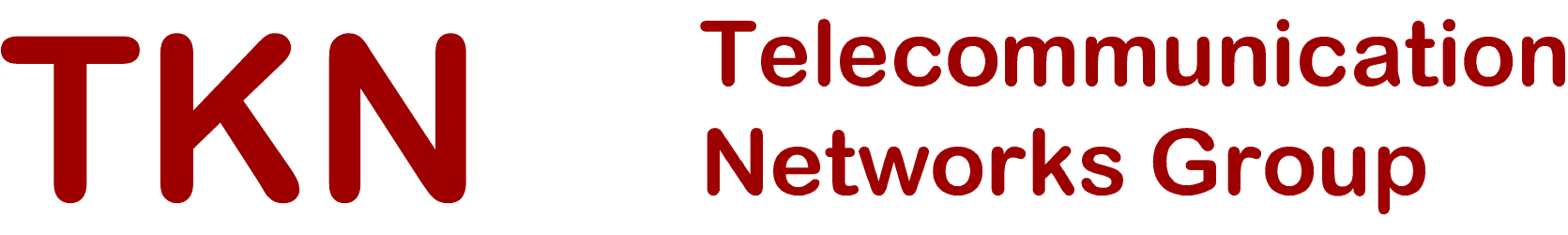}
  \\
\end{tabularx}
\setlength{\tabcolsep}{6pt} % back to default

\vspace{1.0cm}

\begin{center}
{\huge
\noindent
Technische Universität Berlin

\vspace{0.5cm}

\noindent
Telecommunication Networks Group

\begin{center}
\rule{15.5cm}{0.4pt}
\end{center}
}
\end{center}

\begin{minipage}[][11.0cm][c]{14.5cm}
{\Huge

\begin{center}
\trtitle
\end{center}

\begin{center}
{\LARGE \trauthor} \\
{\Large \tremail}
\end{center}

\begin{center}
Berlin, \trdate
\end{center}

\vspace{0.5cm}

}

\begin{center}
\setlength{\fboxrule}{2pt}\setlength{\fboxsep}{2mm}
\fbox{TKN Technical Report \trnumber}
\end{center}

\end{minipage}

\setlength{\fboxrule}{0.4pt}
\setlength{\fboxsep}{0.4pt}

\begin{center}

  \rule{15.5cm}{0.4pt}

  \vspace{0.5cm}

  {\huge {TKN Technical Reports Series}}

  \vspace{0.5cm}

  {\huge Editor: Prof. Dr.-Ing. Adam Wolisz}

  \vspace{0.5cm}

 \end{center}

}

\begin{abstract}

%!TEX root = ../resfi-vap.tex

\begin{abstract}
\subsection*{\abstractname}

The increased usage of IEEE 802.11 Wireless LAN (WLAN) in residential environments by unexperienced users leads to dense, unplanned and chaotic residential WLAN deployments. Often WLAN Access Points (APs) are deployed unprofitable in terms of radio coverage and interference conditions. In many cases the usage of the neighbor's AP would be beneficial as it would provide better radio coverage in some parts of the residential user's apartment. Moreover, the network performance can be dramatically improved by balancing the network load over spatially co-located APs.

We address this problem by presenting \textbf{N}eighborhood e\textbf{x}tensible \textbf{WLAN} (NxWLAN) which enables the secure extension of user's home WLANs through usage of neighboring APs in residential environments with zero configuration efforts and without revealing WPA2 encryption keys to untrusted neighbor APs. NxWLAN makes use of virtualization techniques utilizing neighboring AP by deploying on-demand a \textbf{W}ireless \textbf{T}ermination \textbf{P}oint (WTP) on the neighboring AP and by tunneling encrypted 802.11 traffic to the \textbf{V}irtual \textbf{A}ccess \textbf{P}oint (VAP) residing on the home AP. This allows the client devices to always authenticate against the home AP using the WPA2-PSK passphrase already stored in the device without any additional registration process.

We implemented NxWLAN prototypically using off-the-shelf hardware and open source software. As the OpenFlow is not suited for forwarding native 802.11 frames, we built software switch using P4 language. The performance evaluation in a small 802.11 indoor testbed showed the feasibility of our approach. NxWLAN is provided to the community as open source.

\end{abstract}
\end{abstract}

%\begin{keywords}
%WLAN, Residential WiFi, Virtualization, Software-defined Networks
%\end{keywords}

\tableofcontents

% ================================================================
% Begin here
% ================================================================

%!TEX root = ../resfi-vap.tex

%%%
\chapter{Introduction}

In recent years we have seen a rapid growth of IEEE 802.11 wireless LAN (WLAN) usage in residential or home networks. According to Cisco Visual Networking Index (VNI)~\cite{cisco} by the end of 2019 more than half of the worldwide IP traffic will be generated by WLAN devices. Moreover, according to Watkins et al.~\cite{analytic} already in 2015 nearly 70 percent of all households worldwide that own a broadband Internet access, have their own WLAN Access Point (AP) for providing wireless Internet access and connectivity between household devices. These two trends lead to an ever higher spatial density of AP deployments in urban or sub-ruban residential environments. 

Shi et al.~\cite{shi2015little} evaluated the potential for mutual WiFi sharing in residential environments and revealed that even in sparsely-populated suburban areas mutual WiFi sharing can be advantageous. Unfortunately, today's residential WLAN users cannot gain from AP densification as the wireless access is in general restricted to the user's home AP. Even a simple WLAN sharing requires additional management configuration and above all it cannot provide the same kind of functionality, i.e. although the user has Internet access he cannot access the devices in his home (W)LAN, e.g. printers. Moreover, there are country-specific regulation like the Störerhaftung in Germany~\cite{stoerer} which makes simple WLAN sharing impossible.

\begin{figure}[!ht]%
	\begin{center}
		\includegraphics[width=0.7\linewidth]{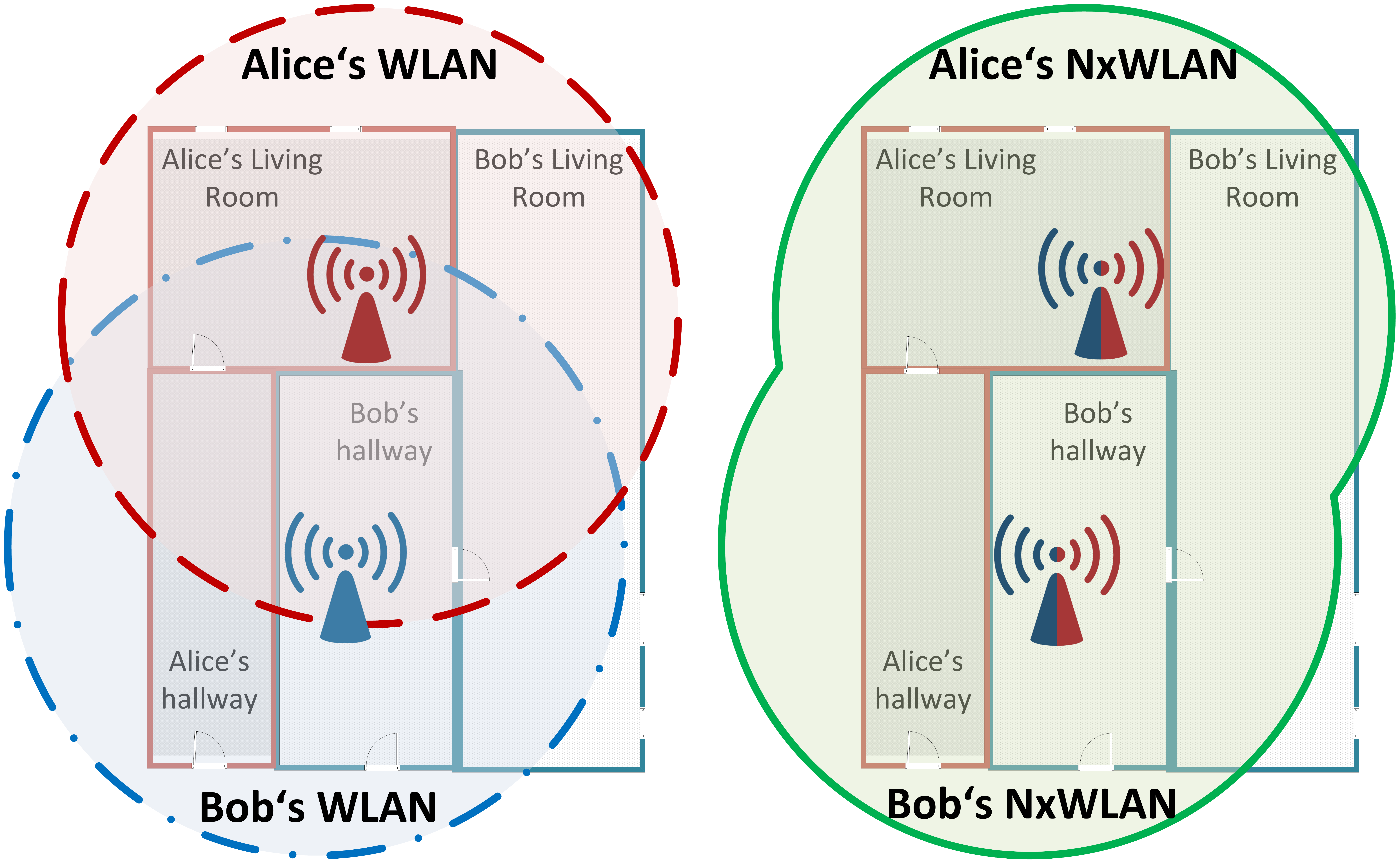}
	\end{center}
	\vspace{-10pt}
	\caption{NxWLAN enables secure extension of users's home WLAN through usage of neighboring residential APs with zero configuration efforts and without revealing WPA encryption keys to unstrusted neighbor APs.}
	\label{fig:motivation}
\end{figure}

In this paper we present NxWLAN (\textbf{N}eighborhood e\textbf{x}tensible \textbf{WLAN}) --- Fig.~\ref{fig:motivation}, which enables secure extension of users's home WLANs through usage of neighboring APs in residential environments with zero configuration efforts and without revealing WPA encryption keys to unstrusted neighbor APs. NxWLAN makes use of virtualization techniques utilizing neighboring AP by deploying on-demand a Wireless Termination Endpoint (WTP) on the neighboring AP and by tunnelling encrypted 802.11 traffic to the virtual AP (VAP) residing on the home AP~\cite{vestin2013cloudmac}. This allows the client devices to always authenticate against the home AP using the WPA-PSK passphrase already stored in the device. There is no need for a registration process or software to be install on the user's devices.

The main contributions of this paper are: i) the design of NxWLAN, an architecture enabling the transparent use of neighboring residential APs to securely access all network devices in the user's home WLAN, ii) a prototypical implementation of NxWLAN using off-the-shelf hardware and open source software like ResFi~\cite{resfi} and P4~\cite{bosshart2014p4} and iii) a performance evaluation in a small 802.11 indoor testbed.
Finally the NxWLAN prototype is provided as open source: https://github.com/nxwlan

%!TEX root = ../resfi-vap.tex

\chapter{ResFi \& SDN Primer}

The proposed solution makes use of two technologies which are described below.

\section{ResFi Framework}

ResFi~\cite{resfi} is a framework enabling the creation of distributed Radio Resource Management (RRM) functionality in residential WLAN deployments. The radio interface of participating APs is used for efficient discovery of adjacent APs and as a side-channel to exchange connection configuration parameters. Those parameters namely the public IP of each AP's RRM unit and security credentials such as symmetric and asymmetric cryptographic keys are then used to build up secured communication tunnels between adjacent APs via the Internet. Different RRM applications like channel assignment and interference management were implemented on top of ResFi.

ResFi is available as open-source and provides a well-defined northbound and southbound API. While the southbound API enables vendors and researchers easily to connect their current AP solution with the ResFi framework, the extensible northbound API is used by ResFi application developers to implement their own RRM solution. The ResFi run-time supports the concurrent execution of multiple applications.

\section{Software-defined Networking}

The Software-defined Networking (SDN) approach was originally designed and used only in wired networks. However, there have been already a lot of attempts to adapt it into wireless and mobile networks~\cite{schulz2014aeroflux}~\cite{yiakoumis2014sdn}~\cite{ali2013crowd}. The SDN paradigm allows for decoupling control and data plane. All forwarding functions (data plane) are implemented in hardware and it is possible to program (control plane) its behavior using protocols like OpenFlow~\cite{mckeown2008openflow}.

Although, the OpenFlow interface gives network operators a simple and flexible approach for implementation of new functionalities in their networks, the number of header fields, that are supported is quite limited, what prevents the more advanced mechanisms to be implemented.
%First version of the OpenFlow was supporting only fields of Ethernet and TCP/IPv4 headers, then gradually support for more advanced protocols, including MPLS, was introduced.

Unfortunately, as the new protocols are being designed and existing ones evolve, it is hard for the OpenFlow consortium to keep up with updating specification. Moreover, since today's OpenFlow protocol is limited to programming flow rules only on Ethernet-based traffic, it is not suited for WLAN networks, i.e. it is not possible to match on 802.11 frames nor use measurements present in RadioTap header (like RSSI value). 

For those reason, a new approach for Software Defined Networking was introduced. The P4~\cite{bosshart2014p4} is high-level language for programming forwarding plane of packet processors. The P4 language is i) protocol-independent, meaning it is not designed for any specific protocol, but it provides a way to express protocol formats in a common syntax; and  ii) target-independent, i.e. the P4 program can be executed across different platforms, including NPUs, FPGAs, software and re-configurable hardware switches.

%!TEX root = ../resfi-vap.tex

\chapter{NxWLAN's Design Requirements \& Principles}

The main goal of NxWLAN is to improve user experience in residential WiFi networks by providing possibility for transparent usage of neighboring APs to extend the coverage of the user's home WLAN. To this end, we have identified a couple of requirements that have to be fulfilled:
\begin{itemize}
  \item Avoiding any additional (re)configuration on WLAN APs and client STAs,
  \item Seamless switching between neighboring APs for mobility and radio coverage issues,
  \item Possibility to steer client STAs to lightly loaded APs like it is possible in Enterprise WLANs for load-balancing issues,
  \item Providing strong E2E encryption and elimination of risk of man-in-the-middle attack,
  \item Providing layer-2 access to all devices in Home (W)LAN, including printers,
  \item Advanced radio resource management, e.g. sophisticated Rf channel assignment, MAC layer tuning.
\end{itemize}

Since residential WiFi networks are setup and managed by people lacking experience and knowledge of wireless networking, we argue that any new solution should not overwhelm users with complicated configuration. Basically, any additional configuration steps have to be limited or, in best case, completely eliminated. Plug-and-Play approach and Self-configuration capabilities should limit interaction with users only to let them configure their WPA-PSK passphrase. 

In dense deployments of residential WiFi networks, it is highly possible that some parts of user's apartment are in radio coverage of neighbor's AP rather than home AP, cf. Fig.~\ref{fig:motivation}. Even locations with coverage to both the home and neighboring AP can gain from AP sharing in case of unequally network load, e.g. use neighboring AP in case the home AP is highly loaded. We would like to allow users to gain from this fact by allowing them to use neighboring APs to transparently access its own WLAN. One possible solution is the on-demand deployment of Virtual AP (VAP) on neighboring APs with the same configuration as in the home AP. The main drawback of this solution is that the entire home AP configuration, including WPA-PSK passphrase, has to be shared and stored in neighboring APs, what creates the possibility for interception (untrusted environment). Another option is to create VAPs with new configurations, but this requires either manual interaction from users or a special application running on client STAs to reconfigure WLAN connection. Finally, the coverage of home WLAN can be increased by deploying a lightweight WTP on the neighboring AP and tunneling encrypted 802.11 frames between the WTP and the VAP in the home AP~\cite{anyfi,vestin2013cloudmac}. This allows the client STA to transparently authenticate with the home AP using WPA-PSK passphrase already configured in the user device. Note, that tunneling encrypted 802.11 frames eliminates the need to share credentials with neighboring APs, what gives strong end-to-end security between home AP and STAs and reduces possibility of man-in-the-middle attack.

Another important issue that we would like to address is to provide clients connected over neighboring AP an access to all devices in their Home (W)LAN (layer 2). We believe that users moving around their apartment would like to still have an access to their printer and network-attached storage (NAS).

As the available capacity at APs varies, the clients should always be connected through the AP that provides the best performance, e.g. high throughput or low latency. Moreover, transition between neighboring AP has to be performed seamlessly without requiring user to reconnect manually. The decision about switching the client to neighboring AP has to take multiple factors into account, namely network load in radio access of home and neighboring AP; and available Internet backbone capacity and latency between those two APs. We have designed an algorithm that takes mentioned factors as input and gives willingness of serving new node as output.

%!TEX root = ../resfi-vap.tex

%%%
\chapter{NxWLAN's Architecture}

\section{Overview}

\begin{figure}[!ht]
   \begin{center}
       \includegraphics[width=1\linewidth]{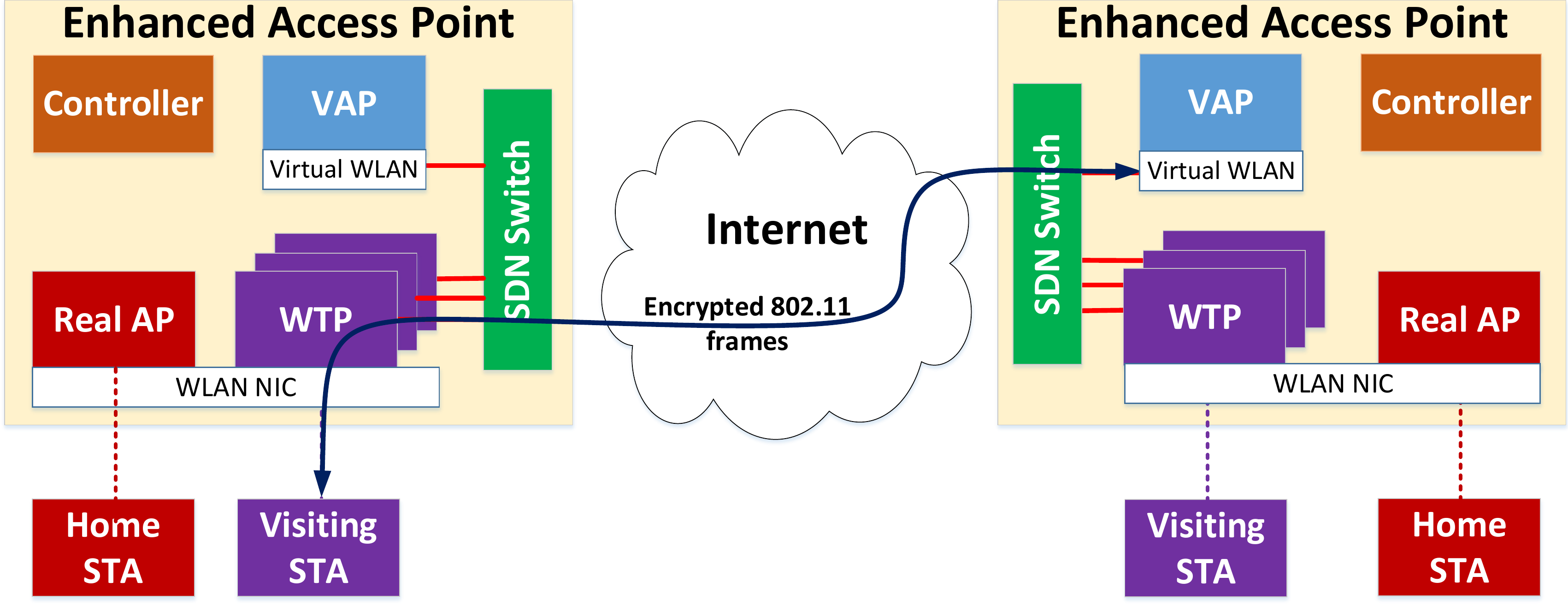}
   \end{center}
    \vspace{-10pt}
   \caption{Architecture of residential NxWLAN network.}
   \label{fig:resfi_arch}
\end{figure}

Fig.~\ref{fig:resfi_arch} shows the overview of the NxWLAN architecture. Each Enhanced Access Point (EAP) consists of a Real (Home) Access Point (RAP), a Virtual Access Point (VAP), set of Wireless Termination Points (WTPs), a SDN switch and a controller. 
The RAP is the legacy AP that is always on and is the only choice for STAs when there are no neighboring APs available. 
The VAP is setup on-demand when discovering NxWLAN-enabled neighboring APs willing to share their wireless and back-haul resources for visiting client stations. It takes care of the generation of all 802.11 management and data frames and about encrypting them if required. The VAP is responsible for authentication and association of STAs that are connected through neighboring APs. Note that the VAP uses the same configuration as the RAP, so a STA can connect to it without any additional configuration. 
The WTP is a simple radio head that is: i) transmitting all encrypted 802.11 frames received from the VAP via the network tunnel and ii) forwarding all received encrypted 802.11 frames to the VAP. Moreover, the WTP is responsible for generation of all 802.11 control frames, especially ACK frames; and lower MAC functions including rate control algorithm. An EAP sets up a WTP for each neighboring NxWLAN-enabled EAP. 
The VAPs and the WTPs are interconnected using network tunnels. Note, that a single VAP can be connected with multiple WTPs and it may serve STAs connected over different WTPs. To this end, the SDN switch is needed to forward traffic received from WTP to the corresponding VAP and vice-versa.

The Controller is creating and managing all entities present in a EAP and communicates with other controllers within neighboring EAPs in a secure way using the Resfi framework. Specifically, we have implemented our controller as a Resfi application~\cite{resfi}. 

\section{WTP/VAP Setup Procedure}

\begin{figure}[!ht]
   \begin{center}
       \includegraphics[width=0.55\linewidth]{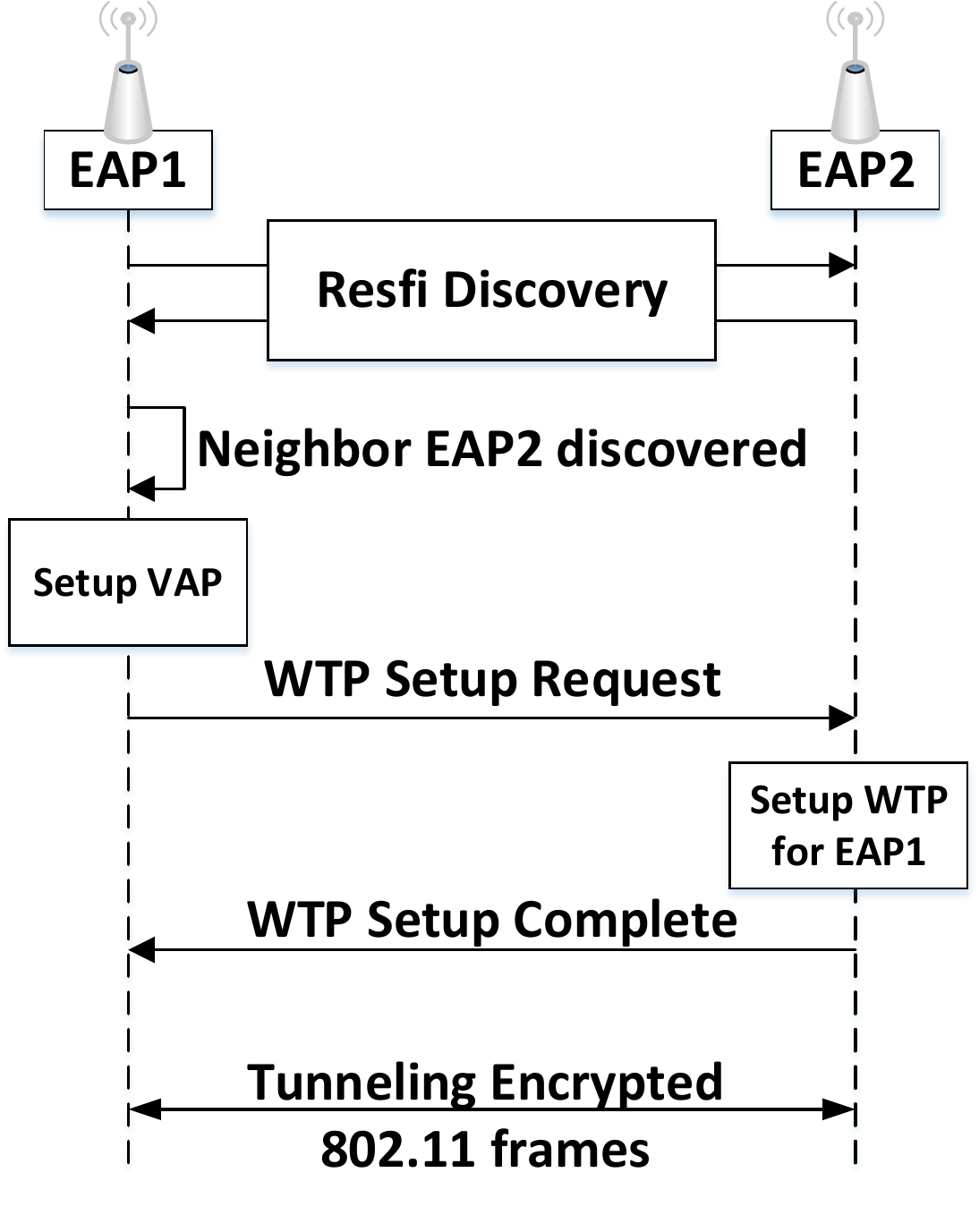}
   \end{center}
    \vspace{-10pt}
   \caption{AP neighbor discovery and setup of WTP and VAP.}
   \label{fig:sequence_diagram_1}
\end{figure}

Fig.~\ref{fig:sequence_diagram_1} illustrates the setup procedure. Upon discovery procedure, the EAP creates a VAP and sends \emph{WTP Setup Request} messages to all discovered neighboring APs. According to set policies, neighboring EAPs may reject or proceed this request. After successful creation of WTP and tunnel, EAP responds with a \emph{WTP Setup Complete} message to the requesting neighbor AP. In case of rejection, it does not respond. Upon reception of confirmation message, EAP completes VAP configuration and sets up a tunnel according to the configuration received from the neighbor. From this point, VAP and WTPs are able to transmit 802.11 frames over tunnel.

For the SDN switch we selected the P4 language. This is because OpenFlow supports only a limited number of header fields and is therefore not able to operate with native 802.11 frames. In CloudMAC~\cite{vestin2013cloudmac}, authors were able to implement a switch for 802.11 frames using an OpenFlow software switch, but frames were encapsulated with an Ethernet header with specific Ethertype value before they enter the switch. By using the P4 language we were able to develop a software switch that is able to switch native 802.11 frames and forward them over tunnels between VAP and WTP. The P4 software switch contains a table that allows the forwarding of the frames to the proper tunnel interface based on MAC destination. By manipulating the switch table, STAs and their traffic can be easily roamed between different WTPs. 

Our system does not require any modification or reconfiguration in client STAs. Basically, STAs are not even aware that they are served by a VAP.

\section{STA Association Procedure}\label{sec:assoc}

\begin{figure}[!ht]
   \begin{center}
       \includegraphics[width=0.8\linewidth]{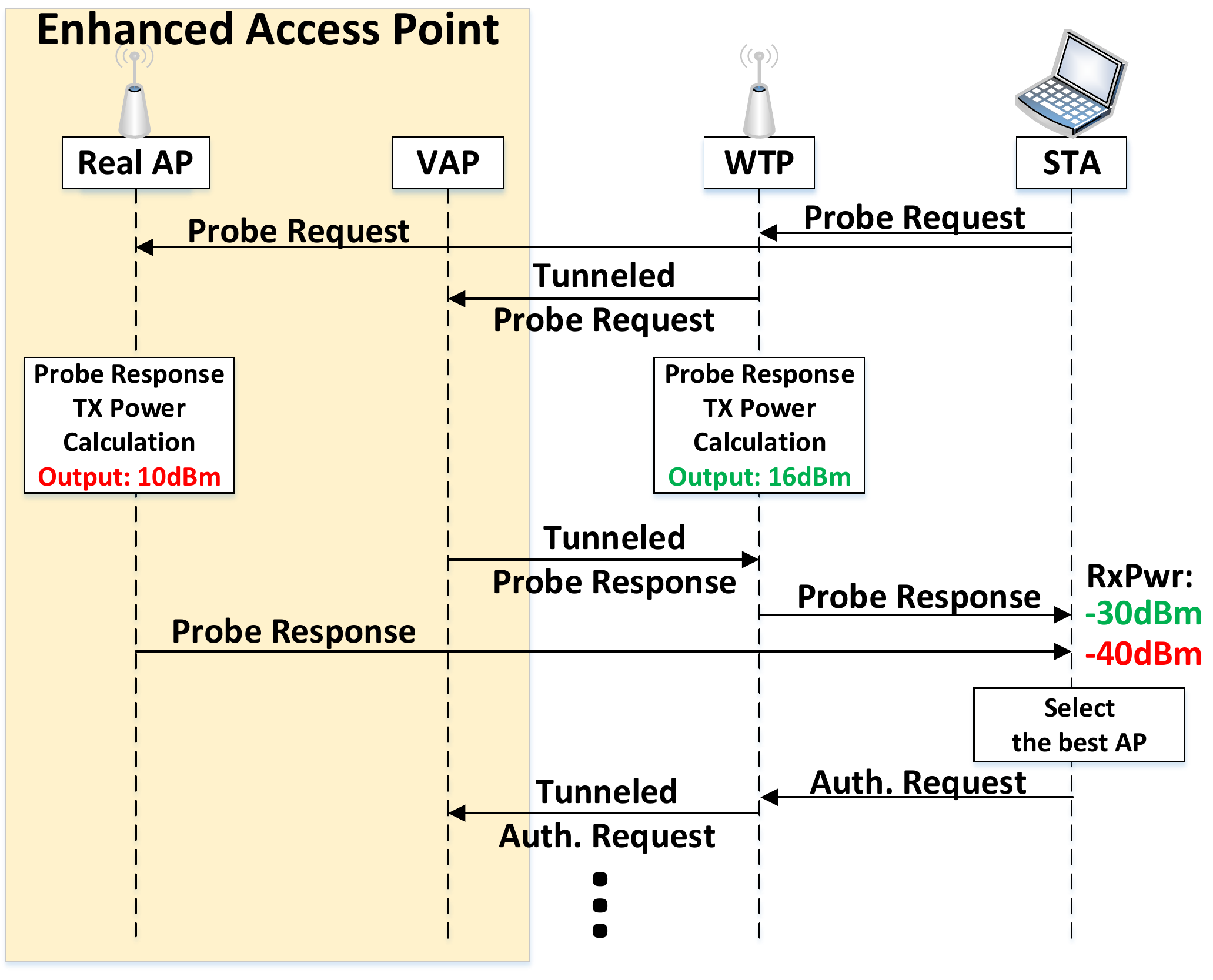}
   \end{center}
    \vspace{-10pt}
   \caption{Client association process in NxWLAN.}
   \label{fig:sequence_diagram_2}
\end{figure}

%\tolja{Piotr: please check the equations and pseudocode!!}
Fig.~\ref{fig:sequence_diagram_2} presents a STA association procedure. In order to connect to an AP, a STA performs an active scan by broadcasting a \emph{Probe Request} sequentially over all available WLAN channels as specified in the 802.11 standard. The \emph{Probe Request} can be received by ether a Real AP directly or by WTPs and tunneled to the VAP. In legacy WLAN, all APs that are in communication range of the STA should send an unicast \emph{Probe Response} immediately upon the reception of a \emph{Probe Request}. 

With NxWLAN especially in dense residential deployments a specific client can be served by different APs - either directly by its home AP (RAP) or by its WTP deployed on a neighboring AP and connected to the VAP on home AP. For load balancing and coverage reasons NxWLAN would like to steer any new client STA to a particular AP based on the AP's willingness to serve the client which depends on its available capacity in the radio and back-haul network. Unfortunately, the Internet-scale latency of the control channel between the APs~\cite{resfi} makes any cooperation between APs impossible as the client stations waits only a short time for the \emph{Probe Response} packet before scanning another Rf channel.

NxWLAN tackles that issue as follows. The APs are able to compute their willingness of serving the new client independently of each other. Moreover, we exploit the fact that STAs always connect to the AP which provides the strongest signal, i.e. highest received power in \emph{Probe Response}. In particular in NxWLAN each AP encodes its willingness to serve the client in the TX power of the \emph{Probe Responses}. If TX power is selected properly, it is possible to steer the client STA to a intended AP. Note, as we always send the \emph{Probe Response} (with possibly different TX power values) our solution does not create a failed scanning, i.e. the client STA is able to associate after a single scan procedure (probing). In our example in the Fig.~\ref{fig:sequence_diagram_2}, both RAP and WTP are operating on the same channel (but they can also use different channels) and receive the same \emph{Probe Request}. Then both entities calculate TX power and send \emph{Probe Responses}. Since RAP is more willing to serve new client, it uses higher TX power. Finally, the STA connects to RAP, as its \emph{Probe Response} was received with the highest RSSI value.

We have designed and implemented an algorithm for computing the TX power, that takes the available capacity in both the wireless and the wired network (backhaul) in account (Listing~\ref{calc_probe_reply_txpower}). In order to steer client STAs to a proper AP, we have to assure that the received power of the \emph{Probe Response} sent by the AP with the larger available capacity is higher than the one of an AP with little available resources. Note, our algorithm requires knowledge of the path-loss between STA and APs. Here we assume the transmission power of STA is set to fixed value, do not change over time and is known to all APs. Those assumptions allow us to compute path-loss. As a future work, we would like to exploit 802.11k and 802.11v standards that allow STA to report to AP the RSSI of received beacon frames. 

\begin{figure}[!ht]%
	\begin{center}
		\includegraphics[width=0.8\linewidth]{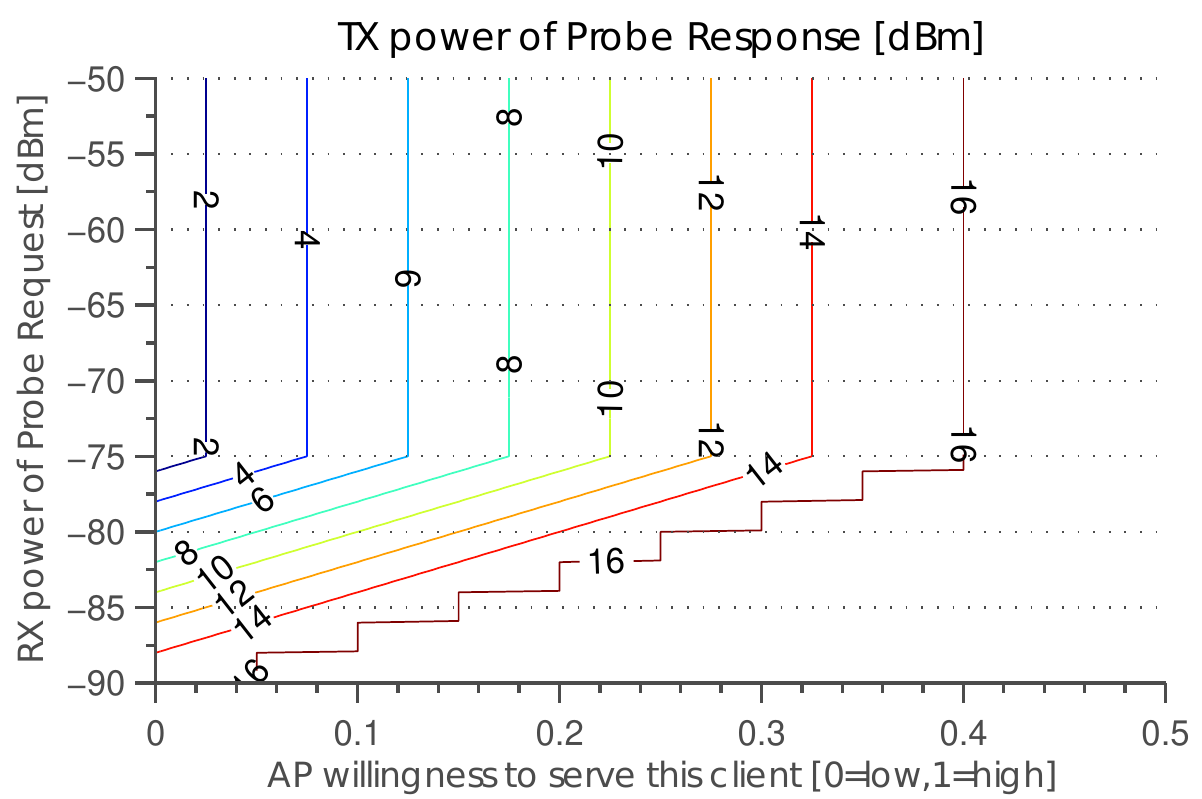}
	\end{center}
	\vspace{-10pt}
	\caption{Impact of AP willingness to serve a client on the transmit power of \emph{Probe Response} packet.}
	\label{fig:probe_reply_txpower}
\end{figure}

The complete algorithm for the calculation of the transmit power for the \emph{Probe Response} packets, i.e. RAP and for each VAP, is depicted in Listing~\ref{calc_probe_reply_txpower}. Note, upon reception of a \emph{Probe Request} this algorithm is executed by each AP independently. Note: we optimize for downlink. The helper function that maps the AP's priority to serve the client to the transmit power of \emph{Probe Response} packet is shown in Listing~\ref{enodeAPWillingness}. Note, this algorithm takes the pathloss to the client into account to make sure that the receive power of the \emph{Probe Response} will be independent of the pathloss aka client distance (Fig.~\ref{fig:probe_reply_txpower}). The function \textit{getMACRate()} calculates the maximum client MAC layer throughput assuming no bottlenecks in the backhaul. When considering a single WLAN BSS and assuming 802.11 DCF packet level fairness and no external interference (WiFi and non-WiFi) the function can be defined as follows:

\begin{align}\label{eq:eff_data_rate}
\textit{getMACRate}() &= \gamma_k \times R_{\mathrm{PHY}}^k
\end{align}
where $R_{\mathrm{PHY}}^k$ is the physical layer rate of client $k$ and $\gamma_k$ is its relative airtime:
\begin{align}\label{eq:rel_airtime}
\gamma_k &= \frac{1/R_{\mathrm{PHY}}^k}{\sum_{i \in C}{1/R_{\mathrm{PHY}}^i}}
\end{align}
where $C$ is the set of active clients in WLAN BSS. Note Eq.~\ref{eq:eff_data_rate} needs to be adapted if you configured 802.11e Transmit Opportunity (TXOP).

\begin{algorithm*}
\scriptsize
\caption{Calculate the transmit power of \emph{Probe Response} packets for Real AP and all VAPs represented by their WTPs.}
\label{calc_probe_reply_txpower}
\begin{algorithmic}[1]
\Require $P_{\mathrm{rx}}^{\mathrm{Preq}}$ \Comment{Receive power of \emph{Probe Request} transmitted by scanning client station.}
\Procedure{calcProbeResponseTxPower}{}
  \State $\mathbb{C} \leftarrow \mathrm{getFatClients()}$ \Comment{Estimate the clients served by this EAP which are generating large network traffic (ResFi).}
	\State $R_{\mathrm{PHY}}^{*} \leftarrow \mathrm{predictPhyRate}(P_{\mathrm{rx}}^{\mathrm{Preq}})$ \Comment{Predict the expected PHY rate of the new client when served by this AP using the measured RX power from \emph{Probe Request} (look-up table).}
	\State $\mathrm{rapDlBh} \leftarrow \mathrm{getAvailableDLBackhaul()}$ \Comment{Estimate available downlink backhaul capacity of AP (ResFi API).}
	\State $\mathrm{rapUlBh} \leftarrow \mathrm{getAvailableULBackhaul()}$ \Comment{Estimate available uplink backhaul capacity of AP (ResFi API).}
	\State $R_{\mathrm{MAC}}^{*} \leftarrow \mathrm{getMACRate}(\mathbb{C}, R_{\mathrm{PHY}}^{*})$ \Comment{Predict expected client capacity at MAC layer when served by Real AP.}
	\State $R_{\mathrm{ALL}}^{*} \leftarrow \mathrm{min}(R_{\mathrm{MAC}}^{*}, \mathrm{rapDlBh})$ \Comment{Take minimum of av. DL backhaul and wireless capacity.}
	\State $p^{*} \leftarrow \mathrm{min}(1, R_{\mathrm{ALL}}^{*} / R_{\mathrm{ALL}}^\mathrm{MAX})$ \Comment{Calculate AP's willingness to serve client (normalized with max rate, e.g. 25 MBit/s): 0-very low, 1-max.}
	\State $P_{\mathrm{tx},\mathrm{rap}}^{\mathrm{Prep}} \leftarrow \mathrm{enodeAPWillingness}(P_{\mathrm{rx}}^{\mathrm{Preq}}, p^{*})$ \Comment{Encode AP's willingness into TX power of \emph{Probe Response}.}
	
 \ForAll{$\mathrm{vap} \in \mathrm{getNeighbors}()$} \Comment{For each neighboring VAP represented by their WTP on this AP (ResFi API).}
	\State $\mathrm{vapDlBh} \leftarrow \mathrm{resfi\_msg}[\mathrm{vap}][\mathrm{'vapDlBh'}]$ \Comment{Take reported available DL back-haul capacity of neighbor AP (ResFi).}
	\State $\mathrm{vapUlBh} \leftarrow \mathrm{resfi\_msg}[\mathrm{vap}][\mathrm{'vapUlBh'}]$ \Comment{Take reported available UL back-haul capacity of neighboring AP.}
	
	\State $R_{\mathrm{ALL}}^{\mathrm{vap}} \leftarrow \mathrm{min}(R_{\mathrm{MAC}}^{*}, \mathrm{rapDlBh})$ \Comment{Take the minimum of available backhaul and wireless capacity.}
	\State $\mathrm{R'}_{\mathrm{ALL}}^{\mathrm{vap}} \leftarrow \mathrm{min}(R_{\mathrm{ALL}}^{\mathrm{vap}}, \mathrm{vapDlBh}, \mathrm{vapUlBh})$ \Comment{Consider the 802.11 frame tunneling from WTP to VAP.}
	\State $p^{\mathrm{vap}} \leftarrow \mathrm{min}(1, \mathrm{R'}_{\mathrm{ALL}}^{\mathrm{vap}} / R_{\mathrm{ALL}}^\mathrm{MAX})$ \Comment{Calculate AP's willingness to serve client.}
	\State $P_{\mathrm{tx},\mathrm{vap}}^{\mathrm{Prep}} \leftarrow \mathrm{enodeAPWillingness}(P_{\mathrm{rx}}^{\mathrm{Preq}}, p^{\mathrm{vap}})$ \Comment{Encode AP's willingness into \emph{Probe Response} TX power.}
 \EndFor\\
 \Return $P_{\mathrm{tx},\mathrm{rap}}^{\mathrm{Prep}}, P_{\mathrm{tx},\mathrm{vap_1}}^{\mathrm{Prep}}, \ldots, P_{\mathrm{tx},\mathrm{vap_N}}^{\mathrm{Prep}}$ \Comment{Return the TX power values for all \emph{Probe replies} (RAP and all VAPs).}
\EndProcedure
\end{algorithmic}
\end{algorithm*}

\begin{algorithm*}
\scriptsize
\caption{Helper function encodes AP's willingness to serve the given client into the transmit power of \emph{Probe Response} packet, i.e. high values means high willingness (e.g. low network load at AP).}
\label{enodeAPWillingness}
\begin{algorithmic}[1]
\Require $P_{\mathrm{rx}}^{\mathrm{Preq}}$ \Comment{Received power (dBm) from client's \emph{Probe Request}.}
\Require $w$ \Comment{AP's willingness to serve this client.}
\Require $P_{\mathrm{tx}}$ \Comment{The default client transmit power (to be known).}
\Require $P_{\mathrm{rx}}^{\mathrm{low}}$ \Comment{The minimum receive power required for reception of \emph{Probe Request}, e.g. -90\,dBm}
\Require $P_{\mathrm{rx}}^{\mathrm{high}}$ \Comment{The max possible receive power of \emph{Probe Request}, e.g. -50\,dBm.}
\Procedure{enodeAPWillingness}{}
  \State $PL \leftarrow P_{\mathrm{tx}} - P_{\mathrm{rx}}^{\mathrm{Preq}}$ \Comment{Pathloss of link AP-client.}
	\State $P_{\mathrm{tx}}^{\mathrm{min}} \leftarrow \mathrm{max}(1,P_{\mathrm{rx}}^{\mathrm{low}} + PL)$ \Comment{Minimum TX power to be used to avoid outage.}
	\State $P_{\mathrm{tx}}^{\mathrm{Prep}} \leftarrow \mathrm{min}(P_{\mathrm{tx}}, P_{\mathrm{tx}}^{\mathrm{min}} + w \times (P_{\mathrm{rx}}^{\mathrm{high}} - P_{\mathrm{rx}}^{\mathrm{low}}))$ \Comment{Assumption: client is sending \emph{Probe Request} with max power.}\\
 \Return $P_{\mathrm{tx}}^{\mathrm{Prep}}$ \Comment{Return \emph{Probe Response} TX power (dBm).}
\EndProcedure
\end{algorithmic}
\end{algorithm*}

%!TEX root = ../resfi-vap.tex

%
%
%
\chapter{NxWLAN's Implementation}

\begin{figure}[!ht]
   \begin{center}
       \includegraphics[width=1\linewidth]{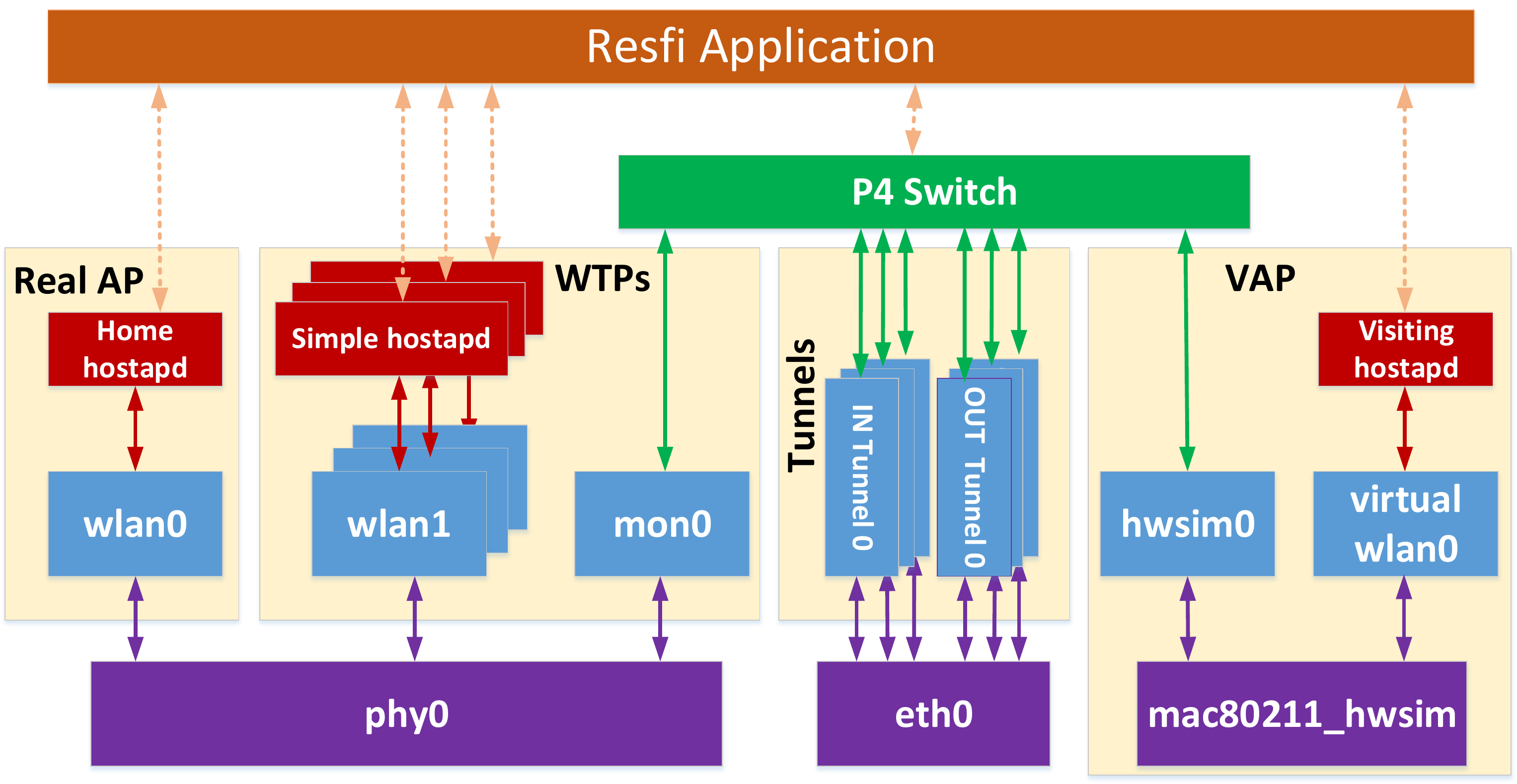}
   \end{center}
    \vspace{-10pt}
   \caption{NxWLAN implementation details - components inside a single residential Enhanced AP.}
   \label{fig:resfi_impl_arch}
\end{figure}

We have implemented the NxWLAN prototype using open-source tools. Fig.~\ref{fig:resfi_impl_arch} shows details of its implementation which are discussed below. 

\section{VAP Rate Control}

Wireless channels are extremely variable and may change their propagation characteristics very fast being affected by many factors, like interference and multi-path fading. Therefore, wireless transmitters need to continuously adjust the bitrate (modulation and coding scheme) used for packet transmission. The most commonly used rate adaptation algorithm for Linux based systems is the Minstrel rate control algorithm~\cite{minstrel}. 

The rate adaptation has to be done quickly to keep up with wireless channel dynamics, i.e. must be faster than channel coherence time. In CloudMAC, the WTP transmits frames with PHY rate present in RadioTap headed and selected by VAP~\cite{VestincentralizedRate}. Authors also introduced API for rewriting the rate using OpenFlow, what gives a possibility of creating centralized rate control algorithm. As frames are tunneled between WTPs and VAPs, what introduces an additional delay, we argue that rates selected in the VAP may be already outdated when they are transmitted through the WTP. For that reason, in NxWLAN the WTPs are responsible for rate adaptation. The challenge is that the Minstrel algorithm does not work for frames injected over monitor interfaces. Those frames are generally transmitted using a fixed (lowest) rate or using a rate specified in RadioTap header.

We observed, that the reason why Minstrel is not working, is that there are no rate-retry tables created for not associated client STAs. Those tables are created during the association phase which is missing on the WTPs, because clients authenticate against the VAP on the home AP. To overcome this problem, we create an additional new virtual wireless interface over the physical one for each WTP and run a simplified version of hostapd on it. This hostapd is not able to generate any frames and was stripped from almost all functionalities, besides STA association. Using the functionality of our previous work to transfer the state of STA from one hostapd instance to another~\cite{Zubow16bigap_seamless_handover}, we are able to associate STAs by calling proper functions exposed by our simplified version of hostapd via the hostapd\_cli command line tool. In this way, we create rate-retry tables for client STAs that are further managed and filled by minstrel.
%Moreover, we modified transmission path of IEEE80211 mac subsystem to query those tables for selected rate also for frames injected over monitor interface.

%
%
%
\section{Control frames}

Every unicast frame transmission has to be acknowledged by the receiver. The 802.11 standard defines that an ACK frame has to be sent after SIFS which equals $10\,\mu$s which makes tunneling it through Internet backhaul between WTP and VAP which introduces delay in order of (tens of) milliseconds infeasible, i.e. transmitter will always assumes that the packet is lost and unnecessarily retransmit the data packet and hence wasting valuable radio resources. For this reason, all low level functions of 802.11 MAC, including acknowledgment and channel access (DCF), has to be realized by the WTP. In general, because of timing requirements, all control frames have to be generated by the wireless interface of the WTPs.

On the other side, the virtual wireless interface in the corresponding VAP, also waits for ACK frames. We have modified the hw\_sim kernel module (that is used to create virtual WLAN NICs) to immediately acknowledge every frame transmission, which is recognized by virtual NIC as perfect radio conditions and makes it transmit always with the highest possible rate. Of course, it may happen that real channel between WTP and STA do not allow for transmission with such rate. In that case, frames generated by VAP has to be enqueued in WTP, which introduces additional delay. 
%\piotr{check: Currently, we do not configure any queueing disciplines in WTP, but use default configuration (pfifo?, dropping?)}. 
We argue that in case of TCP it would not be a problem, because TCP/IP performs flow/congestion control taking the available link capacity into account.
%\sven{I think this is already working because the frames are tunneled somewhow which means there must be a sender and receiver buffer on both sides. And retransmissions will be handeled by minstrel on the WTP, so everything is already working as described right?}

%Due to delay introduced between VAP and WTP and tight time requirements for control frames (PS-POLL and ACK), the support for power saving mode is currently not available. For that reason, we argue that only clients that can afford operation without power saving mode can be served by VAP. Battery-constrained devices, that requires operation in power saving mode, has to be served by Real AP, and currently cannot gain from our solution.

%
%
%
\section{Frame Tunneling}

Tunneling 802.11 frames between the VAP and WTPs is realized over L2TP tunnels, that are created after AP neighbor discovery during bootstrap phase. The L2TP protocol is used for interconnecting LAN networks over Internet. The main drawback of L2TP is that it does not provide encryption of packets. We argue that in our case it is not a problem, because the payload of 802.11 frames is already encrypted even before entering tunnel interface. The 802.11 headers are not encrypted, what in case of eavesdropping, may give intruder the knowledge who is using our solution. Therefore, if additional encryption applied over the whole 802.11 frames is needed, the L2TP solution can simply be replaced by other tunnel solution such as TLS or DTLS.

%Moreover, since NxWLAN always tries to connect clients to best WTP, it creates the possibility of tracking clients in terms of location. 

Usually, two tunnels are created between two neighbors, each for one VAP-WTP pair. A single tunnel carries frames in both directions. It would be sufficient to use only one tunnel, but that would require mixing and filtering frames of different neighbors, what would be more problematic that just creating two tunnels.

Currently, L2TP protocol is used, what make it easy to create and remove tunnels, but on the other hand it requires installing additional dependency in Linux OS. In~\cite{p4_dc}, authors implemented VXLAN tunneling protocol inside P4 switch. As we already use P4 software switch in NxWLAN, we believe tunneling may also be implemented inside it, what would reduce number of required dependencies.

\section{P4-SDN Switch}

P4 introduces an abstract switching model, where switch containing a set of programmable stages that packets travel successively. First, incoming bits are parsed into packets by sequence of parsers. Then, packets enter an ingress pipeline consisting of a sequence of match-action tables, that may modify the packet header before sending it to the next one. The ingress pipeline determines the egress port(s), that is set in packet metadata, and the queue where to send packet. The ingress pipeline may take following actions to a packet: forward, replicate, drop, sent to control plane. Next stage on packet way is Buffering subsystem that is responsible for switching and/or replicating packets to output ports. The Egress Pipeline also consists of a sequence of match-action tables for further packet processing. Finally, packets are sent to output port, where deparser (hidden) serializes them.

\begin{figure}[!ht]
   \begin{center}
       \includegraphics[width=0.6\linewidth]{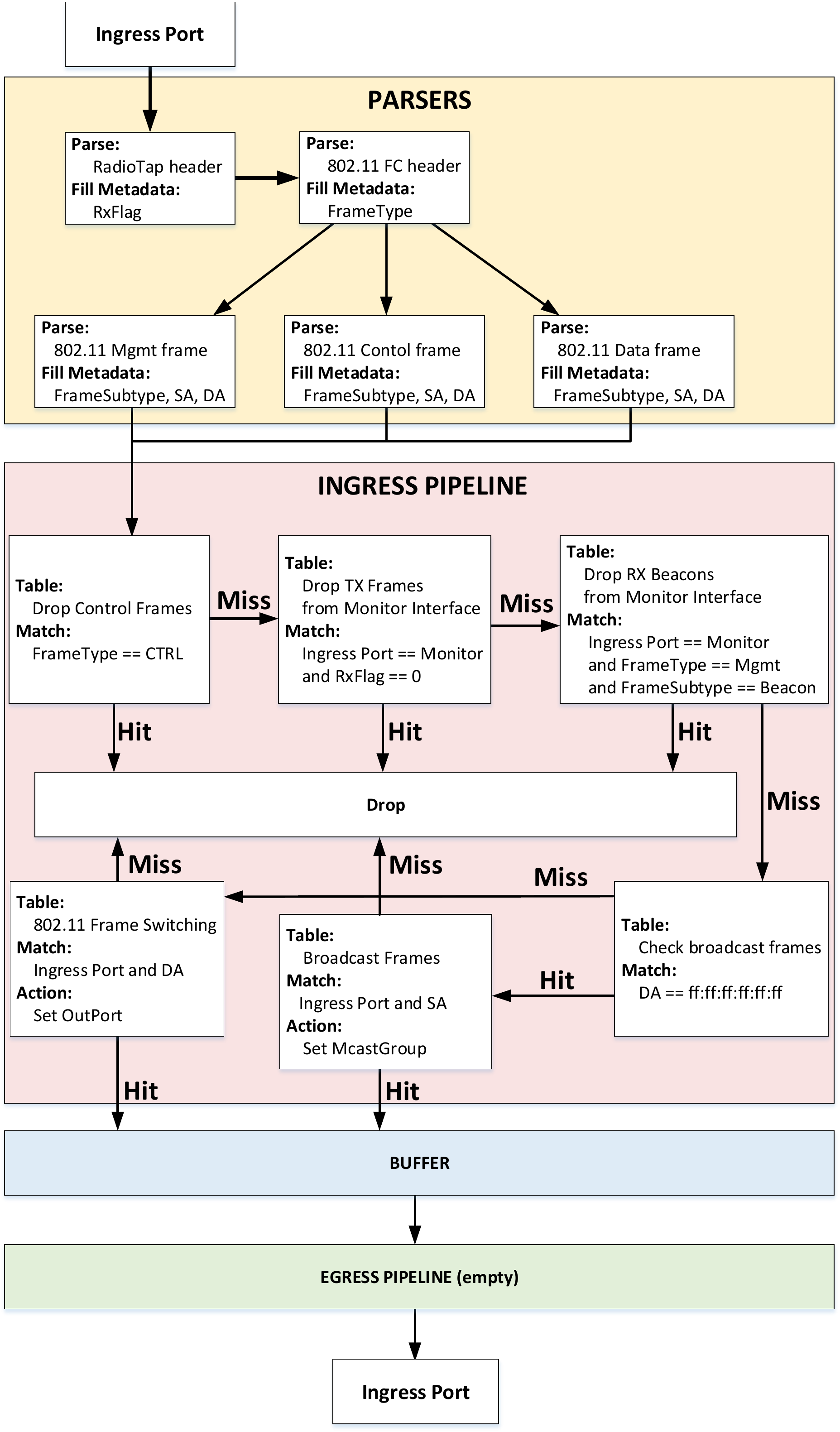}
   \end{center}
    \vspace{-10pt}
   \caption{P4 Switch --- Packet Processing Pipeline}
   \label{fig:p4_pipeline}
\end{figure}

We present internals of packet processing pipelines implemented in P4 language in the Fig.~\ref{fig:p4_pipeline}. We have implemented header definitions of all 802.11 headers and RadioTap headers. Using those definitions, we created parsers sequence for incoming packets. In Ingress Pipeline, we divided tables into two stages. In first stage, we drop unneeded frames. We drop all 802.11 control frames as they are not forwarded to VAP (due to latency) and WTP is responsible for their generation locally. Since successful transmission of frames injected over monitor interface are always reported (for monitoring), we filter them out. In order to save backhaul resources, we drop all \emph{Beacon} that are sniffed on monitor interface. Then packets go to second stage of match+action tables. First, we check if frame has to be switched or broadcasted, based this check packet is sent to one of two programmable tables. The table \emph{802.11 Frame Switching} determines output port, based on available rule entries, ingress port and destination address in 802.11 frame. The table \emph{Broadcast Frames} set multicast group (set of output ports) bases on available rule entries, ingress port and source address. Rule entries and multicast groups are configured by ResFi Control Application.
According to set output the Buffering subsystem decides to: drop packet; forward packet to set output port; or replicate packet over output ports of specified multicast group (in case of broadcast frames). The Egress Pipeline is currently empty. 

\section{Probe Response TX Power Programming}

We modified the ath9k wireless driver to allow the programming of the transmission power of \emph{Probe Response} frames.
%setting the TX power per packet flow. A flow is indentified with flow mark set in metadata associated with packet in Linux OS. 
%In general, we are able to define flows using \emph{iptables}, that allow us define rules for flow classification and mangle mark value in each packet being part of this flow. Then 
%Using the \emph{debugFS} interface, we define transmission power per probe response. 
%The ath9k driver stores list of flows associated with transsmission power. 
Every time, the driver is handling a \emph{Probe Response} frame transmission, it differentiates between \emph{Probe Responses} sent by the RAP or the VAP and takes the appropriate entry in the TX power table filled before via the debugFS interface by the NxWLAN ResFi application. 
%If there is one, tx power in transsmission descriptor structure (metadata that is send to phy along with frame) is overwritten with value defined for flow. Otherwise, driver skip this part and frame is sent with default transmission power that is set to wireless interface.
%Optionally, we are also able to specify the rate that will be used for transmission of frames belonging to this flow. In that case, the rate selected by Minstrel rate adaptation algorithm will be overwritten. 
%\sven{check: As Prope Response packets are not going through Linux netfilter subsystem, nope they are directly sent from hostapd to mac80211 using Netlink, thats why I am not using skbuff mark at all, I will rewrite this section shortly}, it is not possible to use iptables to mark them. For that reason, we have modified \emph{hostapd} to set mark value in skbuff for each Probe Response it is generating. 
The appropriate TX power entries are calculate upon reception of a \emph{Probe Request} by the NxWLAN application, cf. Sec. \ref{sec:assoc}. %In detail calculates willingess of serving new client and encodes it in transmission power that is used during transmission of Probe Response frames.

%
%
%
%\section{Possible Optimizations}
%\sven{I would like to remove this section and maybe merge the major points to future work in conclusion, it makes the paper look like we were too lazy to implement everything correctly}

%As already mentioned, we use monitor interface to inject 802.11 frames. Unfortunately from implementation point, a lot of functionalities like traffic priorization (802.11e EDCA) are missing in monitor interface. \piotr{check: it should be in implmeneted in hardware, so maybe it works? has to check it}. For that reason, we do not support QoS and serve all frames equally. In some cases, it may not be sufficient, for example voice flows should always go over real AP?
%\tolja{discuss traffic priorization: give higher priority to my own traffic; trottle down others? Is this a good idea? I think not.}

%\section{Neighbor VAP Application}
%\tolja{????}
%=======
\section{VAP Bootstrap}

NxWLAN has to be started without any configuration nor manual interaction for each client, that is in areas of good coverage of neighbor AP. In current implementation, tunneling and transmission of beacon frames starts upon discovery and setup of tunnels between two neighboring APs. This might be considered as waste of radio resources, so we working on optimization for NxWLAN to start tunneling 802.11 beacons only when there is at least one STA served by WTP. 

In order to provide seamless roaming worldwide the Anyfi~\cite{anyfi} solution uses a cloud controller for registration and discovery of client station MAC and corresponding home's AP IP address. Whenever the client device comes close to another residential AP the visiting AP lookups the IP address of home AP using the client stations MAC address. In NxWLAN we do not provide roaming, i.e. the NxWLAN is restricted to neighboring APs only. The reason for this is that today all modern devices use MAC address randomization, i.e. disposable interface identifiers, as suggested by Gruteser et al.~\cite{gruteser2005enhancing} in order to improve users’ privacy. As in practice, clients no longer use their real MAC address for \emph{Probe Requests} the Anyfi~\cite{anyfi} approach is unusable.

In NxWLAN we exploit the locality of target environment, i.e. the fact that there is limited number of neighboring APs, and we broadcast \emph{Probe Request} to all VAPs in proximity of the client. This mechanism allows clients using MAC address randomization to gain from usage of NxWLAN.

%\copynpaste{Whenever the device comes close to another residential gateway the embedded Anyfi.net software will detect its presence and send a request to the Controller. The Controller responds with the IP address of the subscriber’s home gateway. The Controller also sends an introduction message to the subscriber’s home gateway to let it the embedded Anyfi.net tunnel termination software prepare for an incoming Wi-Fi over IP connection.}

%For example, a new MAC address can be used for each scan iteration, where one scan iteration consists of sending probe requests on all usable channels. However, since a (draft) specification on MAC address randomization does not yet exist, iOS, Windows, and Linux, all implemented their own variants of MAC address randomization. This raises the question whether their implementations actually guarantee privacy. In the remainder of the paper, we use randomization as a synonym of MAC address randomization.} E.g. \copynpaste{Linux added support for MAC address randomization during network scans in kernel version 3.18. The address should be randomized for each scan iteration [24]. Drivers must be updated to support this feature.}. For the proposed approach MAC address randomization is not an issue because the WTPs are only installed on neighboring APs. Therefore, the Probe Req is simply broadcastet (forwarded) to all VAPs in direct neighborhood.
%>>>>>>> bcdc625c7dc6c3c4d5b06b294c8e29cbc3a46b3c

%!TEX root = ../resfi-vap.tex

\chapter{Evaluation}

The objective of this section is to evaluate the NxWLAN architecture. At first the experimental methodology is introduced, thereafter the results are presented and discussed. 

\section{Methodology}

\begin{figure}[!ht]%
	\begin{center}
		\includegraphics[width=0.5\linewidth]{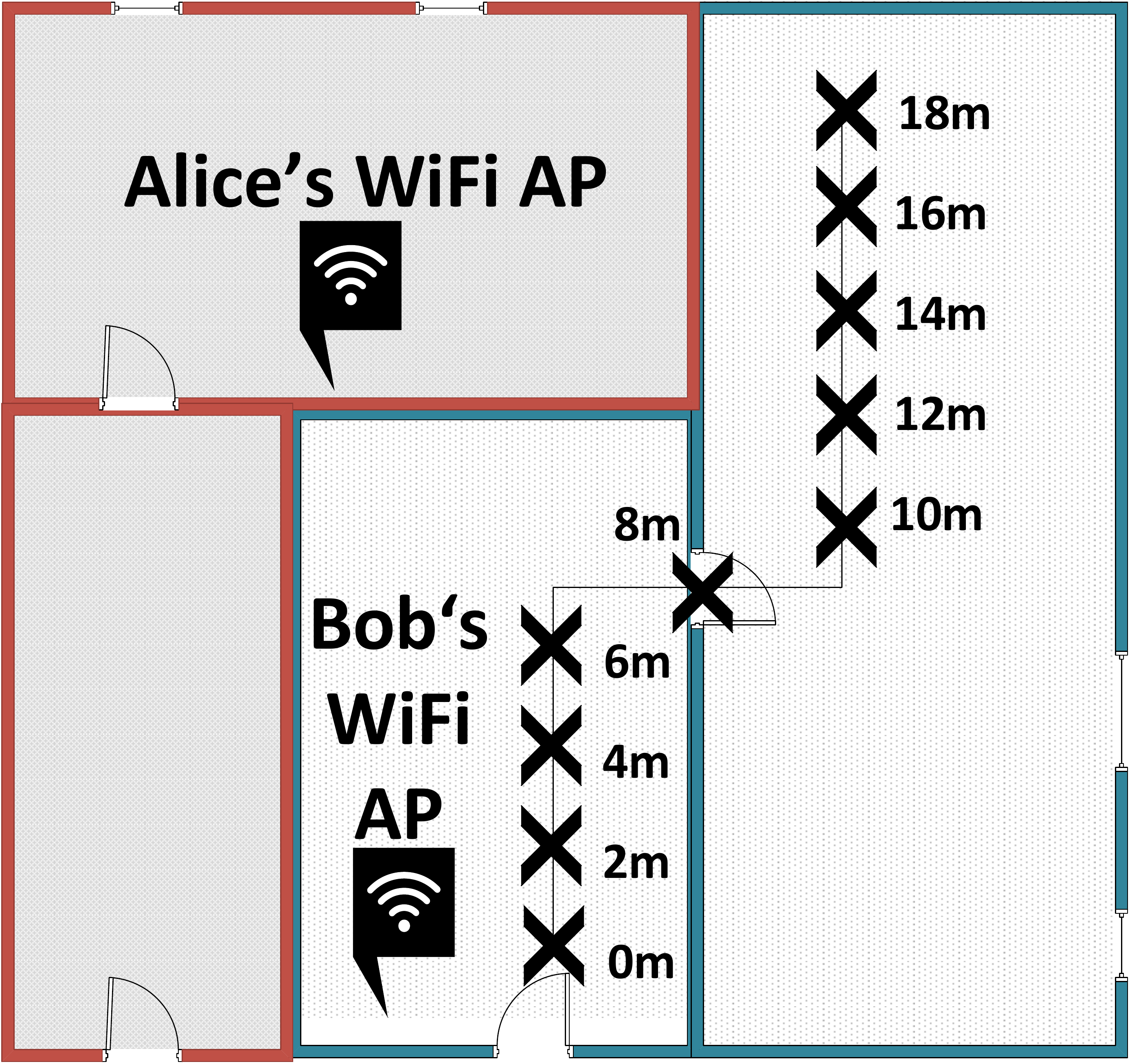}
	\end{center}
	\vspace{-10pt}
	\caption{Experiment setup mimics two adjacent residential apartments each with a single AP.}
	\label{fig:resfi_experiment}
\end{figure}

The proposed NxWLAN is analyzed by means of experiments in a small 802.11n/a testbed. The setup shown in Fig.~\ref{fig:resfi_experiment} mimics two residential apartments each with a single AP. During the experiment a client station was placed at ten different locations within Bob's apartment.

The hardware used for the APs and client stations were standard x86 machines with \textit{Ubuntu 14.04} and \textit{Atheros} WiFi NICs using AR9280 chipsets. For our experiments we set the two APs on two different not otherwise used channels from the 5\,GHz band, i.e. channel 40 and 44. The physical layer was set to 802.11a. As backhaul technology for the residential WiFi deployment we assumed cable modem. Therefore, we used the traffic control tool~\cite{tcUrl} to emulate the last-mile latency in residential WiFi deployments as reported in~\cite{sundaresan2011broadband} in which most users of cable ISPs are in the 0–10 ms interval. As uplink and downlink speed we assumed 50\,Mbit/s.

We considered the following two experiments. First, we analyzed the scenario without any background traffic. Second, in a situation with high background traffic we show that NxWLAN is able to detect and react to the load imbalance by pushing the client to the lightly loaded AP. As performance metric we measured the downlink TCP/IP throughput towards the client station using the iperf tool~\cite{iperf-2013}. We compare NxWLAN in the two mentioned settings with a baseline where the client station can only use its home AP (here Bob).

\section{Results}

\noindent\textbf{Experiment 1: (Extended coverage)} The objective of this experiment is two-fold. First, we want to show that the proposed approach is able to increase the coverage of the home WLAN by using the neighboring AP. Second, that the visiting AP is able to provide a high throughput to the client station although the 802.11 data traffic is tunneled to the home AP. This shows the feasibility of the approach, i.e. rate control and ARQ are working properly on the WTP.

\medskip

\begin{figure}
\begin{minipage}[c][21cm][t]{1\textwidth}
  \vspace*{\fill}
  \centering
  \includegraphics[width=1\linewidth]{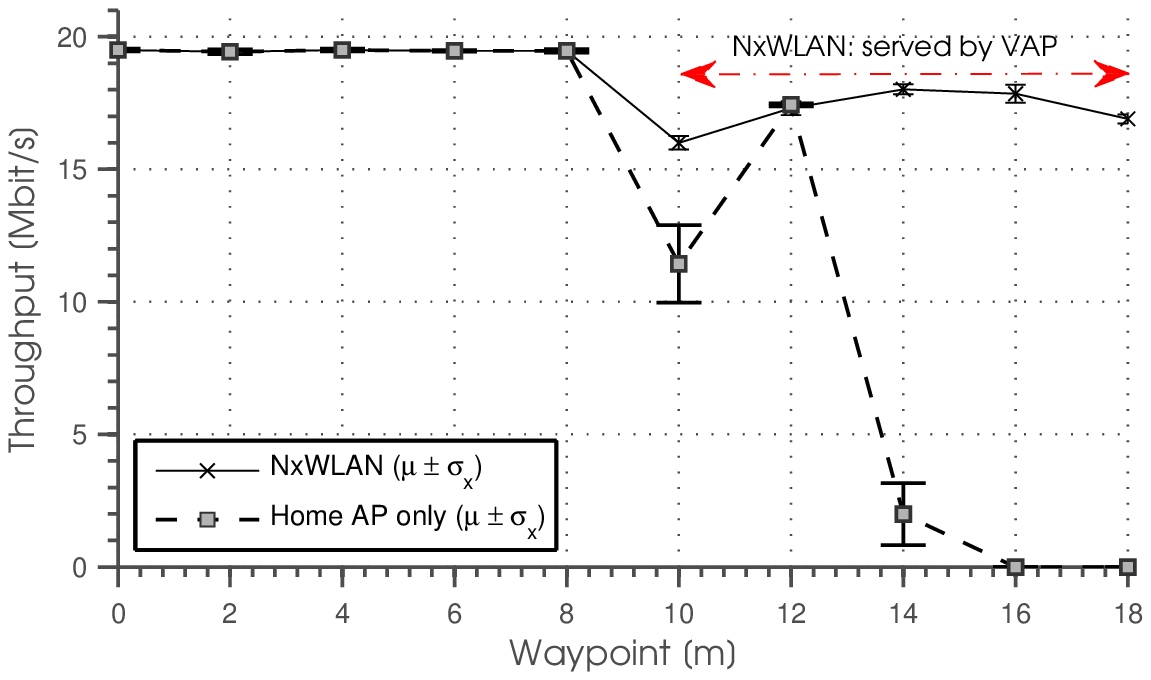}
  \caption{No background traffic: with NxWLAN the client STA is served by WTP in Alice's AP after 12\,m waypoint.}
  \label{fig:resfi_underutilized_channel}\par\vfill
  \includegraphics[width=1\linewidth]{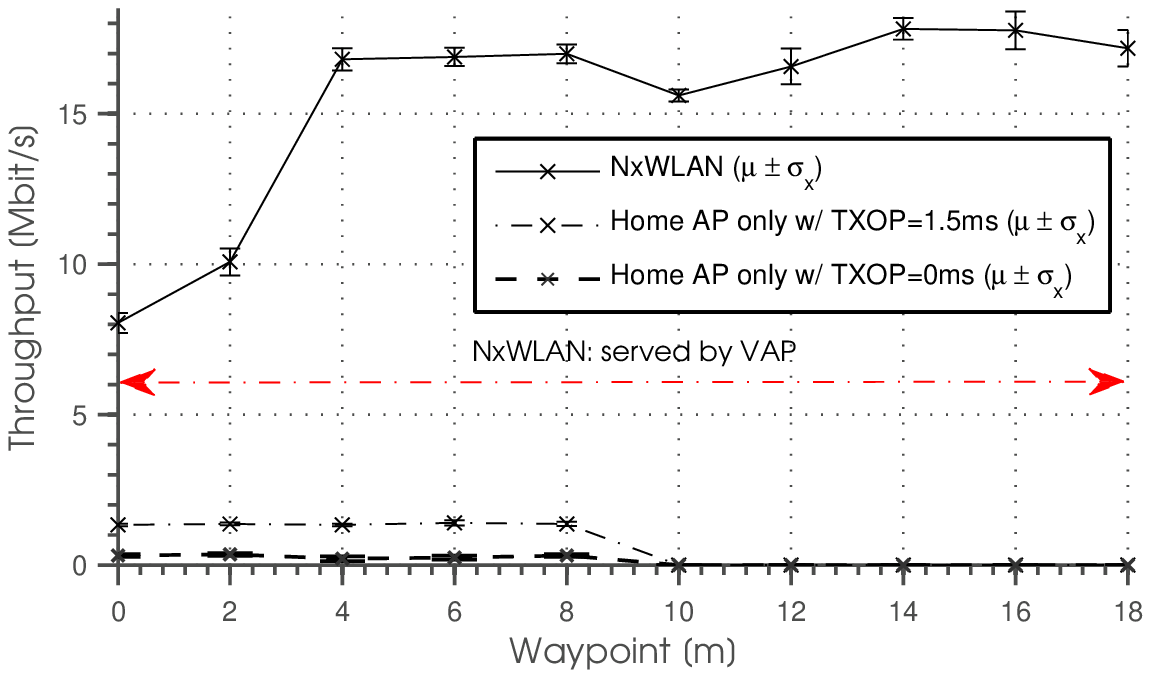}
  \caption{Congested home AP (Bob): with NxWLAN the client STA is always served by WTP in Alice's AP.}
  \label{fig:resfi_congested_channel}
\end{minipage}
\end{figure}

\noindent\textbf{Result 1:} Fig.~\ref{fig:resfi_underutilized_channel} shows the mean and standard error of the downlink throughput to the client station at the ten different locations. Every location point was measured 10 times over a period of 10s. The results reveal that in contrast to baseline NxWLAN is able to provide coverage even at the far corner of the apartment. Moreover, the client achieves a high throughput while being served by the WTP on the neighboring AP (locations 10-18\,m).

\medskip

\noindent\textbf{Experiment 2: (Load balancing)} In this experiment we consider a scenario with unequal network load, i.e. the home AP (Bob) is highly loaded, i.e. two client stations with backlogged TCP uplink traffic emulating cloud storage application, e.g. Dropbox, whereas the neighboring AP (Alice) is unused (idle). The objective is to show that NxWLAN is able to steer the client device to the lightly loaded AP for load balancing reasons by means of manipulating the transmit power of the \emph{Probe Response} packet.

\medskip

\noindent\textbf{Result 2:} Fig.~\ref{fig:resfi_congested_channel} shows again the mean and standard error of the the client downlink throughput at the ten different locations. Again, every location point was measured 10 times over a period of 10s. We can observe that compared to baseline NxWLAN is able to dramatically increase the downlink throughput of the client. This is because in baseline the client station has to share the wireless medium with the two backlogged low-bitrate users (PHY 6 Mbps) in the home AP whereas in NxWLAN the client is steered to the neighboring idle AP (Alice) from the very beginning. Even at very close locations to the home AP our proposed transmit power control for the \emph{Probe Response} packets is able to successfully steer the client device to the far away neighboring AP.

Note, in baseline the performance suffers from uplink/downlink unfairness which can be solved by using Transmit Opportunity (TXOP), here 2\,ms.

%!TEX root = ../resfi-vap.tex

%%%
\chapter{Related Work}

Related work falls into three categories:

\smallskip

\textbf{WiFi sharing for Internet access:} In WiseFi~\cite{shi2015little} authors presented a theoretical concept of an architecture consisting of a smart-phone application and a centralized server that manages the reciprocal WiFi sharing between neighboring APs. Efstathiou et al. 2010~\cite{efstathiou2010controlled} propose a WiFi sharing mechanism to provide city-wide WiFi connectivity for all participating members. Everyone who participates provides free Wi-Fi access to other participants in order to enjoy the same benefit when he is away from his own WLAN network. Participants create their own identities (public-private key pairs) and receive signed digital receipts when they provide Wi-Fi service to other participants. According to their contribution the participants are classified to enforce cooperation and to prohibit free-riders. NxWLAN is different as it does not require any additional configuration/app on the client stations.

\smallskip

\textbf{AP virtualization:} The concept of virtual APs (VAP) managed by a cloud controller was first introduced by Vestin et al. in CloudMAC~\cite{vestin2013cloudmac} to reduce the complexity of management of enterprise WLAN systems. Using VAP and SDN the complexity is moved away from network devices to the cloud. The AP just forwards MAC frames whereas the processing of MAC frames, is implemented on standard servers that are operated in data centers (cloud). NxWLAN is different as it does not require additional cloud infrastructure or central servers, i.e. each EAP hosts the real AP as well as the VAPs. 
Moreover, thanks to usage of P4 switch instead of OpenFlow, the NxWLAN is able to switch native 802.11 frames based on MAC addresses inside them. In CloudMAC frames has to be encapsulated with CloudMAC header before they enter OpenFlow switch. This header has length of Ethernet headed and differs for frames heading towards VAP and WTP.

NxWLAN is related to CAPWAP~\cite{govindan2006objectives} and LWAPP~\cite{calhoun2010lightweight} which also follow a split MAC approach where some MAC frames are generated centrally while others locally by the WLAN card at the AP. However, the split is different in NxWLAN whereas similar to CloudMAC all MAC frames, except for control frames e.g. ACKs, are generated centrally making WTPs stateless components. 

Anyfi~\cite{anyfi} is a commercial product which enables similar user experience as NxWLAN. The client devices simply use the WPA passphrase already stored in the device to authenticate against the home AP (gateway). However, Anyfi differs in the following points: i) it requires a cloud controller for registration and discovery of home APs, ii) it puts the focus in providing seamless worldwide roaming whereas NxWLAN aims to improve the coverage/capacity at home WLAN by using neighboring APs as access, iii) lots of implementation details are unknown like rate adaptation in WTPs and ARQ. One feature of Anyfi, that is worth mentioning, is the Optimizer. It breaks out tunnel between VAP and WTP for Internet-bound traffic and takes care of encryption of client data, thereby it allows for avoiding unnecessary round-trip time. The Optimizer can improve user experience, but on the other hand it requires Temporal Key (TK) from VAP to be shared. In case of Anyfi, the ISP, that is responsible of its deployment, is able to provide secure environment for Optimizer and take needed steps to protect operation of TK transfer. In case of residential networks, it is not possible to provide secure environment, thus Optimizer cannot be applied.

\smallskip

\textbf{Software Defined Networks:} There were already few attempts for introducing SDN concept in wireless network. One available solution for enterprise WLAN network is the Odin~\cite{odin} framework, that is built on top of the OpenFlow controller. The main advantage of Odin is that it provides AP abstraction that simplifies management of clients. It allows network operators to program and deploy enterprise WLAN services as network applications. Furthermore, the abstraction gives application developers a central view on the entire network including all active flows and connected clients. The key component of the framework is the light virtual access point (LVAP), that virtualizes association state and separate them from the physical AP. 
The Odin was designed for enterprise WLAN network with centralized control plane. It consists of a single master, multiple agents and a set of applications that are running in infrastructure. The NxWLAN is a distributed system that is better suited for residential wireless networks.

\chapter{Conclusions}
In this paper we introduced NxWLAN which enables the secure extension of user's home WLANs through usage of neighboring APs in residential environments with zero configuration efforts and without revealing WPA2 encryption keys to untrusted neighbor APs. 

NxWLAN was prototypically implemented on top of the ResFi framework to enable its practical usage in residential WLAN environments. The source code is provided to the community as open source. We evaluated NxWLAN using real COTS hardware in a mimicked residential scenario. The experimental results showed that NxWLAN is able to increase the radio coverage of the home AP while taking the load of the wireless radio channel of both the neighboring and the home AP into account. Moreover, the results show that our approach of adjusting the transmit power of the \emph{Probe Response} frames to steer the client to the desired AP are increasing the achievable throughput of the client STA.  

For future work, we plan to enable the support of power saving for the WTPs by transferring beacon frame generation from the VAP to the WTPs. Moreover, we are working to provide mobility support in NxWLAN using already created handover mechanisms.

Furthermore, we envision that NxWLAN may be generalized into VxWLAN (\textbf{V}irtual e\textbf{x}tensible \textbf{WLAN}). We are working towards roaming support by enhancing NxWLAN to support not only WTP generation on neighbor APs but rather on all APs to which the users client device sends a \emph{Probe Request} searching for its home WLAN network. 

\chapter{Acknowledgment}

This work has been supported by the SDN WiFi Community project funded by Deutsche Telekom.

\bibliographystyle{IEEEtran}
\bibliography{biblio,IEEEabrv}
\end{document}